\documentclass{article}
\usepackage{amsmath,amsthm}
\usepackage{amssymb,latexsym}
\usepackage[mathscr]{eucal}
\usepackage{setspace}
\usepackage{graphics}
\usepackage{array}
\usepackage{nccmath}
\usepackage{geometry}
\usepackage{lineno}
\usepackage[affil-it]{authblk}
\newcommand{\tbf}[1]{\textbf{#1}}
\newcommand{\tit}[1]{\textit{#1}}

\begin{document}
\newgeometry{top = 2in}

\title{\textbf{An Exact Invariant for Relativistic Linear Harmonic Oscillator with Time-dependent Frequency}}
\author{B. K. Shivamoggi\footnote{email: bhimsen.shivamoggi@ucf.edu} and D. Reinhart}
\affil{University of Central Florida \\ Orlando, FL 32816-1364}
\date{}
\maketitle

\begin{abstract}
In this paper we give an exact invariant for a relativistic linear harmonic oscillator with time-dependent frequency. This is accomplished, following Eliezer and Gray \cite{EliezerGray},  for the non-relativistic case, by associating a relativistic plane isotropic harmonic oscillator with this problem. This exact invariant reflects the conservation of angular momentum of the associated relativistic plane isotropic oscillator. Departures in the physical interpretations of this exact invariant caused by relativistic effects are pointed out.
\end{abstract}

\restoregeometry
\newpage

\section{Introduction}
An \tit{exact} invariant for a linear harmonic oscillator with time-dependent frequency was given by Lewis \cite{Lewis}. The existence of conserved quantities implies, according to Noether's Theorem (Lanczos \cite{Lanczos}), symmetry properties of the system in question under certain transformations.  Lutzky \cite{Lutzky} and Shivamoggi and Muilenburg \cite{ShivamoggiMuilenburg} showed that Lewis' exact invariant arises as a consequence of the invariance of the problem under local Lie variational symmetry groups of transformations (Logan \cite{Logan}).

Shivamoggi \cite{Shivamoggi6}, \cite{Shivamoggi7} gave a systematic perturbation procedure to deal with this problem for the case with slowly-varying frequency. Shivamoggi and Muilenburg \cite{ShivamoggiMuilenburg} showed that Lewis' invariant, for the latter case, can be deduced as a power series in the small parameter $\varepsilon$ characterizing the slow variations in the frequency.

An interesting alternative perspective on Lewis' invariant was given by Eliezer and Gray \cite{EliezerGray}. Eliezer and Gray \cite{EliezerGray} pointed out that Lewis' invariant is merely a statement of the conservation of angular momentum of the associated plane isotropic oscillator.

The relativistic harmonic oscillator has been considered by MacColl \cite{MacColl}, Moreau et al. \cite{Moreau}, Li et al. \cite{Li}, among others. The purpose of this paper is to give an exact invariant for the relativistic harmonic oscillator with time-dependent frequency. Because of the difficulty in working with the relativistic version of the local Lie variational symmetry groups of transformations (Logan \cite{Logan2}), we consider the associated higher-dimensional oscillator problem explored by Eliezer and Gray \cite{EliezerGray} previously for the non-relativistic case.

\section{Lewis' Invariant for a Non-relativistic Linear Harmonic Oscillator with Time-dependent Frequency}
Consider a non-relativistic linear harmonic oscillator with time-dependent frequency,
\begin{equation}
\frac{d^2 x}{dt^2}+\omega^2(t)x = 0.
\label{eq::1}
\end{equation}
An exact invariant for this problem was given by Lewis \cite{Lewis},
\begin{equation}
I = \frac{1}{2} \left[ \frac{x^2}{r^2}+ (r \dot{x}-\dot{r}x)^2  \right] = \text{const.}
\label{eq::2}
\end{equation}
where $r(t)$ satisfies the nonlinear differential equation,
\begin{equation}
\ddot{r} + \omega^2(t)r - \frac{1}{r^3} = 0
\label{eq::3}
\end{equation}
and the dots overhead denote derivatives with respect to $t$.

Eliezer and Gray \cite{EliezerGray} pointed out that one can associate a plane isotropic harmonic oscillator with the motion of a non-relativistic linear harmonic oscillator with time-dependent frequency.

One considers a time-dependent central force potential,
\begin{equation}
V = \frac{1}{2} \omega^2(t) r^2
\label{eq::4}
\end{equation}
with the Lagrangian,
\begin{equation}
L = \frac{1}{2} \left[ \dot{r}^2 +r^2 \dot{\theta}^2 -\omega^2(t) r^2 \right]
\tag{5a}
\label{eq::5a}
\end{equation}
or
\begin{equation}
L = \frac{1}{2} \left[ \left( \dot{x}_1 ^2-\omega^2 x_1^2 \right) + \left( \dot{x}_2^2 -\omega^2 x_2^2 \right) \right] \equiv L_1+L_2.
\tag{5b}
\label{eq::5b}
\end{equation}
\setcounter{equation}{5}(\ref{eq::5b}) implies that this problem, in Cartesian coordinates, splits into two independent linear harmonic oscillators of the form,
\begin{equation}
L_i = \frac{1}{2}(\dot{x}_i^2 - \omega^2(t) x_i^2), \hspace{.2in} i = 1,2.
\label{eq::6}
\end{equation}

On noting from (\ref{eq::5a}) that $L$ is cyclic in $\theta$, the azimuthal component of the equation of motion gives the constancy of the angular momentum,
\begin{equation}
h = r^2 \dot{\theta} = \text{const.}
\label{eq::7}
\end{equation}

The radial component of the equation of motion associated with (\ref{eq::5a}) is given by
\begin{equation}
\ddot{r}+r \omega^2-r \dot{\theta}^2 = 0
\label{eq::8}
\end{equation}
and on using (\ref{eq::7}), equation (\ref{eq::8}) becomes
\begin{equation}
\ddot{r}+\omega^2 r- \frac{h^2}{r^3} = 0.
\label{eq::9}
\end{equation}

On noting that,
\begin{equation}
\cos \theta = \frac{x_1}{r}, \hspace{.2in} \sin \theta = \frac{\dot{r}x_1-\dot{x}_1r}{h}
\label{eq::10}
\end{equation}
we obtain
$$
\left( \frac{h x_1}{r} \right)^2 + \left( \dot{r} x_1 - \dot{x}_1 r \right)^2 = h^2 = \text{const.}
$$
or
\begin{equation}
I = \frac{1}{2} \left[ \frac{h^2 x_1^2}{r^2}+ \left(r \dot{x}_1 - x_1 \dot{r} \right)^2 \right] = \text{const.}
\label{eq::11}
\end{equation}
Thus, the exact invariance of $I$ for a linear harmonic oscillator with time-dependent frequency is merely a statement of the conservation of angular momentum of the associated plane isotropic oscillator.

For the case of time-independent frequency, $\omega = \text{const.}$, one solution of equation (\ref{eq::9}) is
\begin{equation}
r = \sqrt{\frac{h}{\omega}} = \text{const.}
\label{eq::12}
\end{equation}
Using (\ref{eq::12}), (\ref{eq::11}) yields
\begin{equation}
I = \frac{h}{2 \omega}(\dot{x}_1^2+ \omega^2 x_1^2) \equiv h \left( \frac{E}{\omega} \right) \equiv h J.
\label{eq::13}
\end{equation}
So, for the case of the standard linear harmonic oscillator, with $ \omega = \text{const.}$, the exact invariant is proportional to the action variable, $J$. Here, $E$ is the total energy of the oscillator.

\section{An Exact Invariant of a Relativistic Linear Harmonic Oscillator with Time-dependent Frequency}

Consider a relativistic linear harmonic oscillator with time-dependent frequency (MacColl \cite{MacColl}, Moreau et al. \cite{Moreau}),
\begin{equation}
\frac{d}{dt}\left[ \frac{\dot{x}}{\left( 1 - \mfrac{\dot{x}^2}{c^2} \right)^{1/2}} \right]+\omega^2(t) x = 0
\label{eq::14}
\end{equation}
where,
$$
\omega^2(t) \equiv \frac{k(t)}{m_0}
$$
$k(t)$ being the time-dependent spring constant, and $m_0$ the rest mass.

 The Lagrangian for the associated relativistic plane isotropic harmonic oscillator\footnote{Consideration of the dynamics in a higher-dimensional space in order to obtain the insights into some results of the corresponding dynamics in a lower-dimensional space is not peculiar to the harmonic oscillator problem. Such an approach goes back to Kaluza \cite{Kaluza}. Kaluza \cite{Kaluza} considered a generalization of Einstein's general relativity from fourth dimensional space-time to fifth-dimensional space-time to achieve the unification of electromagnetism and gravitation.} is given by:
\begin{equation}
\mathscr{L} = c^2 \left( 1- \frac{1}{\gamma} \right) - \frac{1}{2} \omega^2 r^2
\label{eq::15}
\end{equation}
where,
\begin{equation}
\gamma \equiv \left[ 1-\frac{1}{c^2} \left( \dot{r}^2+r^2 \dot{\theta}^2 \right) \right]^{-1/2}.
\label{eq::16}
\end{equation}

Since $\mathscr{L}$ is cyclic in $\theta$, we have for the corresponding canonical momentum,
\begin{equation}
p_{\theta} = \frac{\partial \mathscr{L}}{\partial \dot{\theta}} = \gamma r^2 \dot{\theta} \equiv h = \text{const.}
\label{eq::17}
\end{equation}
from which,
\begin{equation}
\left(r^2 \dot{\theta} \right)^2 = \frac{\mfrac{h^2}{ \gamma_r^2}}{1+\mfrac{h^2}{ r^2 c^2}}
\label{eq::18}
\end{equation}
where,
\begin{equation}
\gamma_r = \left( 1- \frac{\dot{r}^2}{c^2} \right)^{-1/2}.
\label{eq::19}
\end{equation}

The radial component of the equation of motion associated with (\ref{eq::15}) is given by
\begin{equation}
\frac{d}{dt} \left( \gamma \dot{r} \right) - \gamma r \dot{\theta}^2 + \omega^2 r = 0.
\label{eq::20}
\end{equation}
On using (\ref{eq::18}), equation (\ref{eq::20}) becomes
\begin{equation}
\frac{d}{dt} \left[ \left( \gamma_r \sqrt{1+ \frac{h^2}{r^2 c^2}} \right) \dot{r} \right] + \omega^2 r - \frac{h^2}{\gamma_r^2 r^3 \left( 1+ \mfrac{h^2}{r^2 c^2} \right)} = 0.
\label{eq::21}
\end{equation}
On noting that,
\begin{equation}
\begin{aligned}
\cos \theta = \frac{x_1}{r} \\
\sin \theta = \frac{\gamma}{h} \left( \dot{r} x_1 - r \dot{x}_1 \right) = \gamma_r \sqrt{1+\frac{h^2}{r^2 c^2}} \left( \frac{\dot{r} x_1-\dot{x}_1 r}{h} \right)
\label{eq::22}
\end{aligned}
\end{equation}
we obtain
$$
\left( \frac{h x_1}{r} \right)^2 + \gamma_r^2 \left( 1+ \frac{h^2}{r^2 c^2} \right) \left( \dot{r} x_1 - \dot{x}_1 r \right)^2 = h^2 = \text{const.}
$$
or
\begin{equation}
I = \frac{1}{2} \left[ \frac{h^2 x_1^2}{r^2}+ \gamma_r^2 \left( 1+ \frac{h^2}{r^2 c^2} \right) \left( \dot{r} x_1 - \dot{x}_1 r \right)^2 \right] = \text{const.}
\label{eq::23}
\end{equation}
where $r(t)$ satisfies the differential equation (\ref{eq::21}). In the non-relativistic limit, (\ref{eq::21}) and (\ref{eq::23}) reduce to (\ref{eq::9}) and (\ref{eq::11}).

For the case of the time-independent frequency, $\omega = \text{const.}$, one solution of equation (\ref{eq::21}) is
\begin{equation}
r^2 \left( r^2 + \frac{h^2}{c^2} \right) = \frac{h^2}{\omega^2}
\tag{24a}
\label{eq::24a}
\end{equation}
or
\begin{equation}
r^2 = \frac{h}{\omega} \left[ -\frac{h \omega}{2 c^2} + \sqrt{1+\frac{h^2 \omega^2}{4 c^4}} \right] = \text{const.}
\tag{24a}
\label{eq::24b}
\end{equation}
\setcounter{equation}{24}

Using (\ref{eq::24a}),  (\ref{eq::23}) yields
\begin{equation}
I = \frac{h^2}{\omega^2 r^2} \left( \dot{x}_1^2+\omega^2 x_1^2 \right).
\label{eq::25}
\end{equation}

Now, the Hamiltonian for a relativistic linear harmonic oscillator is given by (Li et al. \cite{Li})
\begin{equation}
H = c^2(\gamma-1) + \frac{1}{2} \omega^2 x_1^2
\label{eq::26}
\end{equation}
where,
\begin{equation}
\gamma \equiv \left( 1 - \frac{\dot{x}_1^2}{c^2} \right)^{-1/2}
\tag{27a}
\label{eq::27a}
\end{equation}

Using (\ref{eq::25}), (\ref{eq::26}) becomes
\begin{equation}
H = c^2 (\gamma-1) + \frac{1}{2} \left(\frac{\omega^2 r^2}{h^2} \right) I - \frac{1}{2} \dot{x}_1^2
\tag{28a}
\label{eq::28a}
\end{equation}
from which,
\begin{equation}
I = \frac{h^2}{\omega^2 r^2} \left[ H - c (\gamma -1) + \frac{1}{2} \dot{x}_1^2 \right]
\tag{28b}
\label{eq::28b}
\end{equation}

On noting that from (\ref{eq::27a}) that,
\begin{equation}
\frac{\dot{x}_1^2}{c^2} = 1 - \frac{1}{\gamma^2}
\tag{27b}
\label{eq::27b}
\end{equation}
(\ref{eq::28b}) becomes,
\begin{equation}
I = \frac{h^2}{\omega^2 r^2} \left[ H - \frac{c^2}{2 \gamma^2} (2 \gamma+1) (\gamma-1)^2 \right]
\tag{28c}
\label{eq::28c}
\end{equation}

On using (\ref{eq::24b}), (\ref{eq::28c}) becomes
\begin{equation}
I = \frac{h}{\omega} \left[ \frac{H-\mfrac{c^2}{2 \gamma^2} (2 \gamma+1)(\gamma-1)^2}{-\mfrac{h \omega}{2c^2}+ \sqrt{1+ \mfrac{h^2 \omega^2}{4 c^4}}} \right].
\tag{28d}
\label{eq::28d}
\end{equation}
(\ref{eq::28d}) shows that, in the relativistic case, the exact invariant is no longer proportional to the action variable, $J \equiv H/\omega$, for the standard linear harmonic oscillator, with $\omega = \text{const.}$

\section{Discussion}
In this paper we have given an \tit{exact} invariant for a relativistic linear harmonic oscillator with time-dependent frequency. Because of the difficulty in working with the relativistic version of the local Lie variational symmetry groups of transformations (Logan \cite{Logan2}), we have considered the associated higher-dimensional oscillator explored by Eliezer and Gray \cite{EliezerGray} previously for the non-relativistic case. This exact invariant reflects the conservation of angular momentum of the associated relativistic plane isotropic harmonic oscillator. Departures due to relativistic effects in the physical interpretation of this exact invariant are pointed out.

\end{document}